\documentstyle[11pt,revpasp,epsf]{article}
\begin{document}
\def\etal{\it et al. \rm }

\title{Starburst Galaxies in Clusters}

\author{K.D. Rakos}
\affil{Institute for Astronomy, Univ. of Vienna, Wien, Austria}

\author{J.M. Schombert}
\affil{Dept. of Physics, Univ. of Oregon, Eugene, OR 97403
U.S.A.}

\begin{abstract}

The nature of the starburst phenomenon in galaxies is investigated using a
narrow band color system designed to study color evolution in distant
clusters.  Work on zero redshift, luminous far-IR galaxies, calibrated by
starburst models, demonstrates the usefulness of this color system in
isolating starburst from normal star-forming colors, and also demonstrates
a strong correlation with far-IR colors despite reddening effects.  The
same color system applied to distant clusters finds that a majority of the
faint blue cluster population are starburst dwarf galaxies, probably the
progenitor of the current population of dwarf ellipticals in nearby
clusters.

\end{abstract}

\section{Introduction}

In a series of papers extending over the last 12 years (Rakos \etal 1988,
1990, 1995, 1996, 1997), we have used a narrow band color system to
perform photometry of galaxies in rich clusters at various redshifts in
the rest frame of the cluster. Our study approaches this problem through
the use of a modified Str\"omgren system, modified in the sense that the
filter set is `redshifted' to the cluster of galaxies in consideration;
therefore, no k-corrections. We call our modified system uz,vz,bz,yz to
distinguish it from the original Str\"omgren system (uvby).  The color
indices have been a profitable tool for investigating color evolution and
we have demonstrated that the amplitude of the 4000\AA\ break (Dressier
and Shectman 1987) in the spectra of galaxies is correlated with the
$uz-vz$ color index.

In more recent studies, we have used the high resolution and high S/N
ratio spectra of galaxies published by Gunn and Oke (1975), Yee and Oke
(1978), Kennicutt (1992), De Bruyn and Sargent (1978) and Ashby \etal
(1992) compute synthetic colors. In this fashion, we have constructed a
set of templates to establish the relationship between our color indices
and the morphology of galaxies. This scheme can be expanded upon using
principal component analysis similar to the technique outlined by Lahav
et al. (1996) in order to classification of galaxies in high redshift
clusters.

For our preliminary work, our sample galaxies are divided into four
classes: ellipticals, spirals, Seyferts, and starbursts. Each class is
well separated in our color-color diagrams, especially starburst galaxies
which have anomalous $mz$ indices ($mz$ is defined in the same manner as
the traditional Str\"omgren $m$ indice, $mz=(vz-bz)-(bz-yz)$).  The starburst
galaxies all lie below $mz=-0.2$ due to a combination of intrinsic
reddening combined with starburst colors from a marginally low metal
content population (see Figure 1).  

\begin{figure}
\plotfiddle{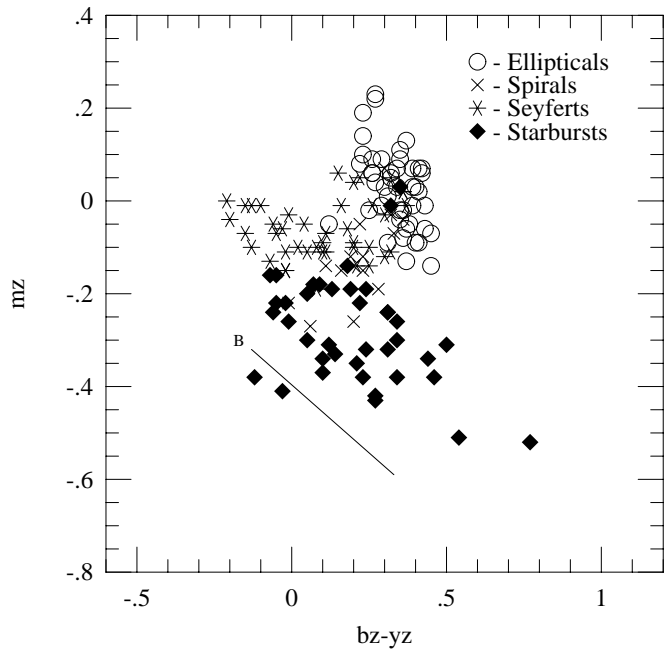}{3.3truein}{0}{70}{70}{-170}{-120} 
\caption{The $mz$ versus $bz-yz$ diagram.  The `B' marks the
position of an BOIa star and the vector is the reddening effect for an
absorption of 3 magnitudes.} \end{figure}

\section{Starburst Colors}

In our photometric system, there are reddening free expressions, similar
to the original relations defined by Str\"omgren, which can be used to
estimate the variable amount of reddening in individual galaxies.  The
reddening free parameter for $mz$, called $C$, is defined to be:

\begin{equation}
C = mz + 0.39 (bz-yz)
\end{equation}

\noindent For $uz-vz$ we define the reddening free variable, $A$, as:

\begin{equation}
A = 1.49 (uz-vz) - (bz-yz)
\end{equation}

Using these variables, our indices become primarily dependent on the age
and total mass of the starburst.  Calibration is based on theoretical
models of starburst galaxies from Leitherer and Heckman (1995).  Figure 2
displays the various models for timesteps of log t = 6.3, 7.0, 7.3. 7.7
and 8.7 years. The initial model at each timestep is for a pure starburst
population.  Additional models represent a the mixture of an underlying
old stellar population that is 1 mag fainter than the starburst, the same
total luminosity as the starburst and 1 mag brighter than the starburst as
noted in the Figure. Note that the luminosity, rather than burst mass, is
used for normalization. The brightness of the burst is taken as the peak
luminosity at 5500\AA.

As an illustration to the use of these indices, we have compared new data
on Mrk325, a clumpy irregular galaxy, and NGC 3277, a normal spiral. Mrk
325 is estimated to contain more than 20,000 very hot stars and is one of
the strongest extragalactic far-infrared sources (see Condon and Yin
1990).  Its luminosity, size and internal motions are all larger than the
typical dwarf irregular galaxies. The position of the two galaxies in the
reddening free diagram are shown as open symbols in Figure 2 and indicate
that the age of the starburst in Mrk 325 is in the range of 40 Myrs,
whereas NGC 3277 displays all the global colors of an old starburst of 0.5
Gyrs ago.

\begin{figure}
\plotfiddle{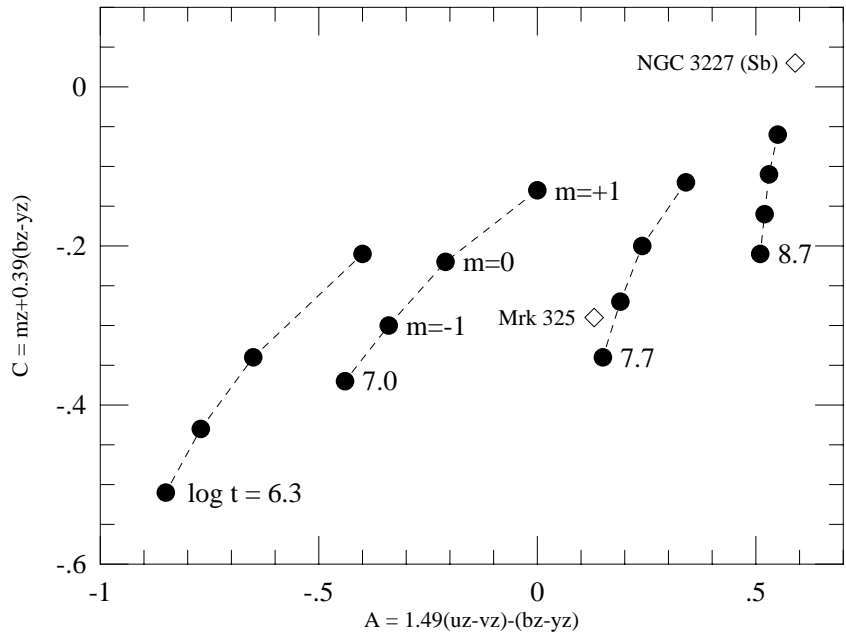}{3.3truein}{0}{70}{70}{-210}{-120} 
\caption{ The age of the starburst in the reddening free color-color
diagram.  The models of Leitherer and Heckman (1995) are shown for various
timesteps of log t = 6.3, 7.0, 7.3. 7.7 and 8.7 years. The initial model
is for a pure starburst population.  Additional models represent a the
mixture of an underlying old stellar population that is 1 mag fainter than
the starburst, the same total luminosity as the starburst and 1 mag
brighter than the starburst.} \end{figure}

The IRAS infrared index $f(60)/f(100)$ and the narrow band colors are also
well correlated despite reddening effects.  The IRAS index is a good
indicator of the nature of the dust heating sources in galaxies and
provides a direct indication of the dominance of the warm component of the
interstellar radiation field produced by O and B Stars.  We have defined a
reddening free color index, $E$ such that:

\begin{equation}
E = (uz-vz) + (vz-yz) - 2.84(bz-yz) 
\end{equation}

Figure 3 shows this correlation between the far-IR colors for starburst
galaxies from the literature and new narrow band data.  It has been
commonly assumed that IRAS colors are poorly correlated to UV and blue
optical colors due to the heavy presence of dust responsible for far-IR
emission. However, new HST imaging has revealed that, in fact,
ultraluminous infrared galaxies display a complex structure of dust lanes
and compact knots of star formation (see Surace \etal 1998).  A
significant amount of the light from these blue star formation regions
exists outside the dust-rich core regions to produce the $E$, far-IR
relation in Figure 3.

\begin{figure}
\plotfiddle{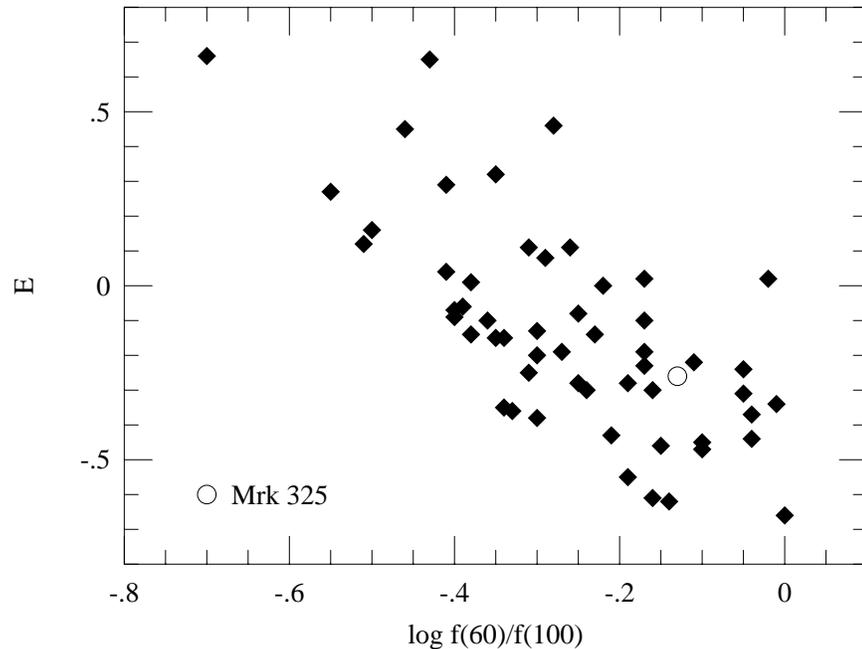}{3.3truein}{0}{70}{70}{-210}{-120} 
\caption{ Narrow band indice $E$ versus the far-infrared color,
$f(60)/f(100)$. The open symbol is the aperture color for the starburst
galaxy Mrk 325.  } \end{figure}

\section{Starbursts in Distant Clusters}

Observations of galaxy clusters with high redshift show an increasing
numbers of blue galaxies ($bz-yz < 0.2$) and an increasing number of those
blue galaxies have $mz < -0.2$ with redshift (Rakos and Schombert 1995).
One such clusters, CL0317+1521 at $z=0.583$, has over 60\% of its
population as blue galaxies and 42\% have $mz < -0.2$, the photometric
signature for starburst.  The deep rest frame Str\"omgren color photometry
of the cluster A2317 ($z=0.211$, Rakos, Odell and Schombert 1997) shows
that the ratio of blue to red galaxies has a strong dependence on absolute
magnitude such that blue galaxies dominate the very brightest and very
faintest galaxies, shown in the bottom panel of Figure 4.  Similar
behavior is found in new data on A2283 ($z=0.183$) also shown in Figure
4.  However, the fraction of galaxies displaying the signatures of a
starburst only increases towards the faint, dwarf end of the luminosity
function (see top panel of Figure 4).

Tidal interactions are frequently invoked as an explanation for the high
fraction of starburst galaxies.  These starburst systems would have their
origin as gas-rich dwarf galaxies who then undergoing a short, but
intense, tidally induced starburst. It should be noted, however, that the
orbits of cluster galaxies are primarily radial, and the typical
velocities into the dense cluster core are high. This makes any encounter
with another galaxy extremely short-lived, with little impulse being
transferred as is required to shock the incumbent molecular clouds into a
nuclear starburst.

Recently, a new mechanism for cluster-induced star formation has been
proposed. This method, called galaxy harassment (Moore \etal 1996)
emphasizes the influence of the cluster tidal field and the more powerful
impulse encounters with massive central galaxies. These two processes
conspire to not only raise the luminosity of cluster dwarfs, but also to
increase their visibility (i.e. surface brightness) and hence their
detectability. One of the predictions of galaxy harassment is that
galaxies in the cores of clusters will be older (post-starburst) than
galaxies at the edges. In terms of star formation history, this is exactly
what has demonstrated in A2317 and A2283 (Rakos, Odell and Schombert
1997).  That is, the blue population is primarily located in the outer
two-thirds of the cluster (see Figure 6 of Rakos, Odell and Schombert
1997).

\begin{figure}
\plotfiddle{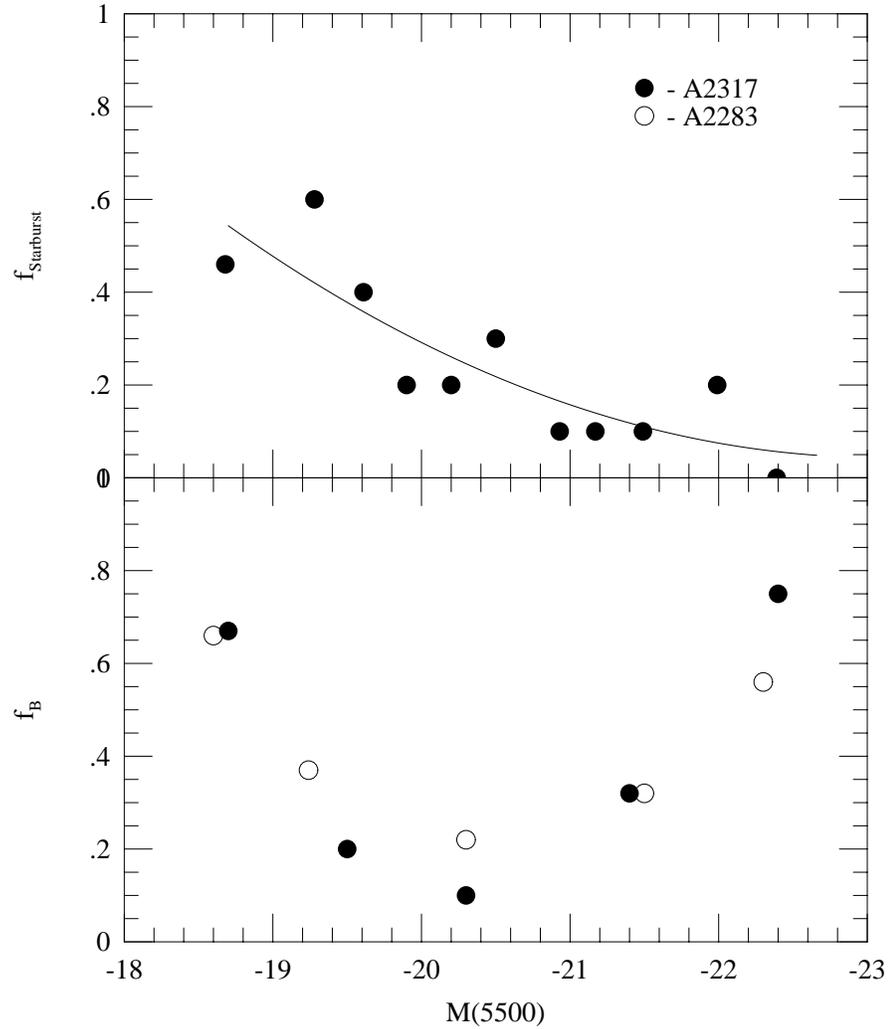}{5.0truein}{0}{70}{70}{-210}{-120} 
\caption{
The fraction of blue galaxies in A2317 ($z=0.211$) and A2283 ($z=0.183$)
as a function of absolute magnitude, $M(5500)$ and the fraction
of starburst galaxies as a function of apparent magnitude in A2317.
} \end{figure}

Regardless of the origin of the blue population, its fate is obvious.
These galaxies do not exist in present-day clusters and, therefore, must
either be destroyed or reduced to the luminosity (or detectability) of
dwarf galaxies.  As shown in Figure 4, the blue galaxies dominate the
bright and faint ends of the luminosity function. Figure 4 also shows that
the blue-fraction of faint galaxies contains a larger number of
starbursting galaxies (i.e. ones with $mz < -0.2$) then the bright
galaxies. One interpretation is that bright galaxies finished their
starburst phase much earlier in the past, and now only display a steady,
spiral-like production of new stars.  Thus, the scenario proposed here is
that the blue galaxies on the bright end of the luminosity function are
core galaxies on low orbits involved in high impulse star-forming events.
Faint galaxies, on the other hand, are cluster halo objects undergoing
harassment style starburst phenomenon.  This scenario naturally divides
the galaxy population by mass and by distance from the center of the
cluster through dynamical processes.  Confirmation for this scenario is
found from deep HST observations by Koo \etal (1997), Oemler \etal (1997)
and Couch \etal (1998) who have shown that the most spectacular starbursts
and emission-line galaxies tend to be low mass objects, whose final state
is likely to be that of a dwarf galaxy.

\section{Conclusions}

The colors of faint blue cluster galaxies in clusters are consistent with
a simple starburst phenomenon and indicates that there exists a bursting
population of dwarf galaxies in clusters which rises in visibility at
earlier epochs, then fades to become the current population of dwarf
elliptical and nucleated galaxies. This becomes a parallel issue to the
faint blue galaxy problem in the field, except in clusters the bursting
dwarf population does not distinguish itself sharply from other cluster
galaxies by color alone.  Only through a mixture of filter indices does
the reddened nature of the bursting dwarf population reveal itself as
unique in color and luminosity from the Butcher-Oemler population common
in most distant clusters.  The fraction of blue galaxies (in A2283 and
A2317) increases on both ends of the luminosity function.  The bright end
is dominated by post-starburst and merger objects identified in HST images
of the Butcher-Oemler population.  The faint end is dominated by the dwarf
population, temporarily enhanced in visibility probably caused by galaxy
harassment mechanisms.

\end{document}